\documentclass[twocolumn,aps,showpacs,groupedaddress,superscriptaddress,prd,floatfix,nofootinbib]{revtex4}

\usepackage{graphicx}
\usepackage{dcolumn}
\usepackage{amsmath}

\usepackage{mathbbold}
\usepackage{amsfonts}
\usepackage{mathrsfs}
\usepackage{graphicx}
\usepackage{amsmath}
\usepackage{amssymb}
\usepackage{bm}
\usepackage{bbm}

\newcommand{\nn}{\nonumber}

\newcommand{\beq}{\begin{equation}}
\newcommand{\eeq}{\end{equation}}
\newcommand{\bqa}{\begin{eqnarray}}
\newcommand{\eqa}{\end{eqnarray}}

\begin{document}
%\preprint{}
%\title{New insight towards hindered magnetic dipole transition in quarkonium
\title{``Hard-scattering" approach to very hindered magnetic-dipole transitions in quarkonium}
%%%%%%%%%%%%%%%%%%%%%%%%%%%%%%%%%%%%%%%%%%%%%%%%%%%%%%%%%%%%%%%%%%%%%%%%%%%%%%
%%%%%%%%%%%%%%%%%%%%%%%%%%%%%%%%%%%%%%%%%%%%%%%%%%%%%%%%%%%%%%%%%%%%%%%%%%%%%%
\author{Yu Jia\footnote{E-mail: jiay@ihep.ac.cn}}
\affiliation{Institute of High Energy Physics, Chinese Academy of
Sciences, Beijing 100049, China\vspace{0. cm}}
\affiliation{Theoretical Physics Center for Science Facilities,
Chinese Academy of Sciences, Beijing 100049, China\vspace{0. cm}}

\author{Jia Xu\footnote{E-mail: xuj@ihep.ac.cn}}
\affiliation{Institute of High Energy Physics, Chinese Academy of
Sciences, Beijing 100049, China\vspace{0. cm}}

\author{Juan Zhang\footnote{E-mail: juanzhang@ihep.ac.cn}}
\affiliation{Institute of High Energy Physics, Chinese Academy of
Sciences, Beijing 100049, China\vspace{0. cm}}
\affiliation{Institute of Theoretical Physics, Shanxi University,
Taiyuan, Shanxi 030006, China\vspace{0. cm}}

\date{\today}

%%%%%%%%%%%%%%%%%%%%%%%%%%%%%%%%%%%%%%%%%%%%%%%%%%%%%%%%%%%%%%%%%%%%%

\begin{abstract}
For a class of hindered magnetic dipole ($M1$) transition processes,
such as $\Upsilon(3S)\to \eta_b+\gamma$ (the discovery channel of
the $\eta_b$ meson), the emitted photon is rather energetic so that
the traditional approaches based on multipole expansion may be
invalidated. We propose that a ``hard-scattering" picture, somewhat
analogous to the pion electromagnetic form factor at large momentum
transfer, may be more plausible to describe such types of transition
processes. We work out a simple factorization formula at lowest
order in the strong coupling constant, which involves convolution of the
Schr\"{o}dinger wave functions of quarkonia with a perturbatively
calculable part induced by exchange of one semihard gluon between
quark and antiquark. This formula, without any freely adjustable
parameters, is found to agree with the measured rate of
$\Upsilon(3S)\to \eta_b+\gamma$ rather well, and can also
reasonably account for other recently measured hindered $M1$
transition rates. The branching fractions of $\Upsilon(4S)\to
\eta_b^{(\prime)}+\gamma$ are also predicted.
\end{abstract}

%%%%%%%%%%%%%%%%%%%%%%%%%%%%%%%%%%%%%%%%%%%%%%%%%%%%%%%%%%%%%%%%%%%%%%%%%%%%%%
% 12.38.-t   Quantum chromodynamics
% 12.38.Bx   Perturbative calculations
% 12.39.Pn   Potential models
% 13.40.Gp   Electromagnetic form factors
% 14.40.Gx   Mesons with S=C=B=0, mass > 2.5 GeV (including quarkonia)

%%%%%%%%%%%%%%%%%%%%%%%%%%%%%%%%%%%%%%%%%%%%%%%%%%%%%%%%%%%%%%%%%%%%%%%%%%%%%%
\maketitle

As a century-old subject, electromagnetic (EM) transitions have been
extensively studied in the fields of atomic, nuclear and elementary
particle physics. EM transition is of considerable interests in
heavy quarkonium physics from both experimental and theoretical
aspects~\cite{Brambilla:2004wf}. Experimentally, it proves to be a
powerful tool to discover new quarkonium states that cannot be
directly produced in $e^+e^-$ annihilation into a virtual photon. A
very recent example is that the long-sought bottomonium ground
state, the $\eta_b$ meson, was finally seen by \textsc{Babar}
collaboration in the magnetic dipole ($M1$) transition process
$\Upsilon(3S)\to \eta_b\gamma$~\cite{Aubert:2008vj}. Theoretically,
it provides a useful means to probe the internal structure, and the
interplay between different dynamic scales in quarkonium.

The standard textbook treatment of EM transitions is based on the concept of
{\it multipole
expansion}~\footnote{The literal meaning of this term is to expand
the electromagnetic field  $A^\mu(t,\mathbf{R}\pm {\mathbf{r}\over
2})$ around ${\bf R}$, the center-of-mass coordinate of
$Q\overline{Q}$ pair, in powers of the relative coordinate $\bf r$,
and the expansion parameter is essentially $kr$, $k$ denoting the photon momentum.
Some authors prefer to dubbing it as {\it long wave-length approximation}.
These two terms are equivalent in this work.}, by assuming the emitted photon to be {\it ultrasoft},
{\it i.e.} $k^\mu \sim m v^2$, where $m$ is heavy quark mass, $v$
denotes the typical velocity of the quark inside a quarkonium.
Consequently, the long wave length of photon cannot resolve the geometrical
details of quarkonium. Obviously, the multipole expansion method
is valid provided that $k r \ll 1$, where $r \sim 1/mv$ is the typical radius of a quarkonium.

One of the great theoretical undertakings is to understand EM
transitions in a situation where the multipole expansion may break down.
It is difficult to find such a situation in atomic system, since the typical
atomic energy spacings are always of order $mv^2\sim m \alpha^2$ (where $\alpha\approx 1/137$
is the fine structure constant).
By contrast, in the realm of QCD, the linearly-rising inter-quark strong force can
host rather highly excited quarkonium states, thus one may encounter EM transitions
in quarkonium with energetic photon.
The aim of this work is to offer a new perspective to tackle such situation.
For definiteness, in this work we will concentrate on the {\it
hindered } magnetic-dipole ($M1$) transition ({\it i.e.}, two quarkonium states with the same
orbital angular momentum but with different spin and principal quantum numbers).
Such study is of practical importance, because it will help one to better understand the
process $\Upsilon(3S)\to \eta_b\gamma$, where the photon carries a
momentum as large as 1 GeV and multipole expansion may cease to
be a decent method.

One usually assumes that the $M1$ transition can proceed without
gluon exchange between $Q$ and $\overline{Q}$. In the
nonrelativistic limit, the transition rate between two $S$-wave
quarkonia is usually described by the well-known
formula~\cite{Brambilla:2004wf}:
%------------------------
\bqa
%------------------------
& & \Gamma[n\,{}^3S_1\to n^\prime\,{}^1S_0+\gamma]
%------------------------
\nn \\
%------------------------
& = & {4\over 3}\, \alpha e_Q^2 \, {k^3\over m^2}
\left|\int_0^\infty \!\!\! dr\,r^2 \,R^*_{n'0}(r) j_0\left({k r\over
2}\right) R_{n0}(r)\right|^2,
%------------------------
\label{M1:LO:formula}
%------------------------
\eqa
%------------------------
where $e_Q$ is the fractional electric charge of $Q$,
$k$ is the photon momentum viewed in
the rest frame of $n\,{}^3S_1$ state, and $R_{nl}(r)$ stands for the
radial Schr\"{o}dinger wave function of quarkonium of principal quantum number $n$
and orbital angular momentum $l$. The spherical
Bessel function $j_0\left({k r\over 2}\right)$ ($j_0(x)\equiv {\sin
x\over x}$) takes into account the so-called finite-size effect (equivalently,
resumming multipole-exanded magnetic amplitude to all orders).
When $k$ is expected to be ultrasoft, it is then legitimate to expand this
function, and the leading contribution to the hindered transition
vanishes due to orthogonality of wave functions. For several
observed hindered transition processes, Eq.~(\ref{M1:LO:formula})
usually yields predictions a few times smaller than the measured
values.

It is widely believed that hindered $M1$ transitions are very
sensitive to the relativistic corrections to (\ref{M1:LO:formula}).
Unfortunately, the way of implementing relativistic corrections
seems to be rather model-dependent. For example, some authors
proposed that, among the contributions from the relativistic
corrections to the $M1$ transition, the hypothesized scalar-part of
the confinement potential may play an eminent role, as well as the
{\it large} anomalous magnetic-dipole moment that may be acquired by
the bound quark due to some nonperturbative
mechanism~\cite{Brambilla:2004wf}. However, both of these
suggestions seem not to be based on a firm and indisputable footing. As a
matter of fact, a variety of quite different predictions to the
transition rates of $\Upsilon(3S, 2S)\to \eta_b\gamma$ have been
made by different authors over the years~\cite{Godfrey:2001eb}. When
confronting with the recently established experimental results,
however, most of them seem not to be favored.

Recently, relativistic corrections to the $M1$ transition have been
readdressed from the angle of nonrelativistic effective field
theories (EFT)~\cite{Brambilla:2005zw}, which allows one to
critically examine the validity of some popular, yet maybe \emph{ad
hoc}, assumptions in many potential model approaches. However, it
still remains a great challenge to accommodate the hindered $M1$
transition in this EFT framework. For instance, after including all
types of conceivable relativistic corrections, the predicted rate
for $\Upsilon(2S)\to \eta_b\gamma$ seems to be much larger than the
measured one.

Impressive progress has been made in calculating the transition rate
of $J/\psi\to \eta_c\gamma$ directly from lattice QCD
simulation~\cite{Dudek:2006ej}. However, it is challenging to analyze
very hindered EM transitions, since excited
quarkonium states will be difficult to probe by lattice simulation.
There is also attempt to model the coupled channel effects for
$\psi^\prime\to \eta_c\gamma$, but such framework seems not to be very predictive
due to existence of several purely phenomenological parameters~\cite{Li:2007xr}.

% We also note that, it seems also difficult for other theoretical
% approaches to account for the hindered $M1$ transitions in a clear way.
%For example, a light-front model
%calculation~\cite{Weiwang:lightfrontmodel} and a light-cone sum rule
%analysis~\cite{YMwang:lcsumrule} give predictions orders of
%magnitude larger than the measured rate for $\Upsilon(3S)\to\eta_b\gamma$.

\begin{figure}[thb]
\begin{center}
\includegraphics[height=3.2cm]{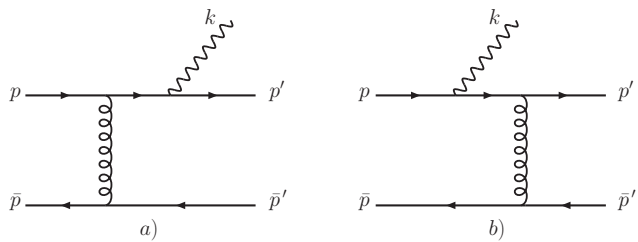}  % moderate label size
\caption{Two of four lowest-order diagrams contributing to hindered
$M1$ transition in our ``hard-scattering" picture.
\label{feynman:diag:gamma}}
\end{center}
\end{figure}

In view of shortcomings of the traditional approaches, we present a
new attempt to analyze very hindered $M1$ transition processes typified by
$\Upsilon(3S)\to \eta_b\gamma$. The key observation is very simple:
in such situation, it is more appropriate to count the radiated photon as
{\it semihard} ($k^\mu \sim mv$, often called {\it soft} in NRQCD
terminology), rather than {\it ultrasoft}. As a consequence,
we should and must give up the notion of multipole expansion.
We further make a key assumption: the leading contribution to such a very
hindered transition is described by Fig.~\ref{feynman:diag:gamma}.
The underlying rationale
is that, in order for the spectator antiquark to join the final
quarkonium state with a significant probability, a {\it semihard}
gluon must be exchanged between $Q$ and $\overline{Q}$ to exert a
kick on it.

It may be worth digressing into pion EM form factor temporarily.
At large momentum transfer, there exists a well-known factorization theorem
for this case~\cite{Lepage:1980fj}:
%------------------------
\bqa
%------------------------
F_\pi(Q^2) &=& \! \int_0^1  \!\!\! \int_0^1\!\! dx dy \,\phi_\pi(x)
T(x,y,Q) \phi_\pi(y)+\cdots,
%------------------------
\label{pion:EM:FF:factorization}
%------------------------
\eqa
%------------------------
where $\phi_\pi$ implies the nonperturbative light-cone distribution
amplitude of a pion, and $T$ refers to the hard-scattering part,
which can be computed in perturbation theory. The lowest-order
contribution to $T$ is also depicted by
Fig.~\ref{feynman:diag:gamma}. It is generally believed that, at
large $Q^2$, this hard-scattering picture is physically more
plausible than the so-called Feynman mechanism (without exchange of
a hard gluon).

We plan to derive a factorization formula analogous to
(\ref{pion:EM:FF:factorization}). In our process, the analogous
``hard-scattering" part is obtained by integrating out the {\it
semihard} mode. We will assume this part is also perturbatively
calculable, crucially because $mv\gg \Lambda_{\rm QCD}$, which seems
legitimate for the $\Upsilon$, presumably even for the $\psi$
family. Quite naturally, we expect that the counterpart of
$\phi_\pi(x)$ in our nonrelativistic problem will be the
Schr\"{o}dinger wave function of quarkonium.

In passing we highlight the very different role played by the {\it
semihard} mode in this work and in Ref.~\cite{Brambilla:2005zw}. In
the latter case when photon is treated as ultrasoft,
the semihard mode can only appear in loop. In contrast to the {\it potential} mode ($p^0 \sim
mv^2,\:\mathbf{p}\sim mv$), it does not make contribution when
descending from NRQCD onto potential NRQCD~\cite{Brambilla:2005zw}.
According to our scheme, however, the semihard mode already makes
crucial contribution at tree level. It is the very mode that we
attempt to integrate out perturbatively, in order to fulfill the
intended factorization.

This said, let us turn to the derivation of the very hindered $^3S_1\to
{}^1S_0$ radiative transition rate. We will perform the calculation
in a covariant fashion at the level of QCD. Since the {\it hard}
($p^\mu \sim m$) quanta decouple in this process, it is also
feasible, perhaps more illuminating, to directly start from NRQCD.
We first note that parity and Lorentz invariance constrain the transition
amplitude to be the form
%------------------------
\bqa
%------------------------
& & \mathcal{M}[n\,^3S_1(P)\to n^\prime\,^1S_0(P^\prime)+\gamma(k)]
%------------------------
\nn \\
%------------------------
 &=&
\mathscr{A} \,\epsilon_{\mu\nu\alpha\beta} \,P^\mu \,
\varepsilon^\nu_{[n\,{^3S_1}]} \, k^{\alpha} \,
\varepsilon_\gamma^{\ast\beta},
%------------------------
\eqa
%------------------------
where ${\varepsilon}_{[n\,{}^3S_1]}$ and $\varepsilon_\gamma$
represent the polarization vectors of the initial quarkonium and the
photon, respectively.
%------------------------
%The relativistic normalization for each particle state has been assumed.
%------------------------
At the rest frame of the initial state, as we will always work in,
the Lorentz structure becomes $\bm{\varepsilon}_{[n\,{^3S_1}]}\cdot
\mathbf{k}\times {\bm \varepsilon}_\gamma^*$, clearly corresponds to
the $M1$ transition. The scalar coefficient $\mathscr{A}$ encodes
all the nontrivial dynamics, and we will proceed to deduce its
explicit form.

We begin with the parton process $Q(p)\overline{Q}({\bar p})\to
Q(p^\prime)\overline{Q}({\bar p}^\prime)+\gamma(k)$, as indicated in
Fig.~\ref{feynman:diag:gamma}. We assign the momentum carried by
each constituent as
%------------------------
\bqa
%------------------------
p &=& {P\over 2}+ q, \qquad\qquad \bar{p}= {P\over2}-q;
%------------------------
\nn \\
%------------------------
p^{\,\prime} &=& {P^\prime \over 2}+ q^\prime,\qquad\quad\;\;
\bar{p}^{\,\prime} = {P^\prime \over 2}-q^\prime, \nn
%------------------------
\eqa
%------------------------
where $q$ and $q^\prime$ are relative momenta inside each pair,
which satisfy $P\cdot q= P^\prime\cdot q^\prime=0$. The invariant
mass of the pairs are $P^2=4 E_q^2$ and $P^{\prime\,2}=4
E_{q^\prime}^2$, and the Lorentz scalars $E_q=\sqrt{m^2-q^2}$,
$E_{q^\prime}=\sqrt{m^2-q^{\prime\,2}}$, which guarantees that each (anti)quark
stays on their mass-shell. Note in the rest frame of
$P^{(\prime)}$, $q^{(\prime)}$ becomes purely space-like.

The quark propagator in Fig.~\ref{feynman:diag:gamma}a) can be
expanded:
%------------------------
\bqa
%------------------------
& & {1\over (p^{\,\prime} + k)^2 - m^2} =  {1 \over k\cdot P^\prime
+ 2 k\cdot q^{\,\prime}}
%------------------------
\nn \\
%------------------------
& \approx &  { 1 \over k\cdot P} +  { 2\,\mathbf{k}\cdot
\mathbf{q}^\prime
 \over (k\cdot P)^2}+\cdots,
%------------------------
\label{quark:propagator}
%------------------------
\eqa
%------------------------
because $\mathbf{k}\cdot \mathbf{q}^\prime \sim m^2 v^2 \ll k\cdot
P^\prime= k\cdot P\sim m^2 v$. We have neglected the small
$q^{\prime\,0}$ component induced by the recoiling of $P^\prime$, as
well as the Lorentz boost effect on $\mathbf{q}^\prime$, which are
higher order corrections. The quark propagator in
Fig.~\ref{feynman:diag:gamma}b) can be expanded in a similar
fashion. Note this expansion is also legitimate when $k$ is
ultrasoft.

Fig.~\ref{feynman:diag:gamma}a) and b) share a common gluon
propagator:
%------------------------
\bqa
%------------------------
{1\over ({k\over 2}+ q^\prime-q)^2+ i\epsilon}
%&=& {1\over k\cdot
%(q^\prime-q)+(q^\prime-q)^2+ i\epsilon}
%------------------------
%\nn \\
%------------------------
& \approx & \! {-1\over (\mathbf{q}^\prime-\mathbf{q})^2
+\mathbf{k}\cdot (\mathbf{q}^\prime -\mathbf{q})- i\epsilon}.
%------------------------
\label{gluon:propagator}
%------------------------
\eqa
%------------------------
Here we retain the $i\epsilon$ term explicitly, for the momentum
integration to be properly evaluated. The two terms in the
denominator are of comparable size, so (\ref{gluon:propagator})
cannot be further expanded. If $k$ is nevertheless counted as
ultrasoft, the second term can be treated as a perturbation. Note
our situation is in drastic contrast to the ordinary NRQCD
calculation for hard exclusive processes. In that case, there is
always a hard scale $\geq m$ in the propagators, so it is safe to
neglect $q^{(\prime)}$ at the zeroth order of NRQCD expansion.

Having specified the concrete forms of the quark and gluon propagators,
we then project the quark amplitude for $Q(p)\overline{Q}({\bar p})\to
Q(p^\prime)\overline{Q}({\bar p}^\prime)+\gamma(k)$ onto the corresponding
color-singlet quarkonium Fock states, with the aid of the covariant
spin projectors accurate to all orders in
$q^{(\prime)}$~\cite{Bodwin:2002hg}, and include the respective
momentum-space wave function for each quarkonium state ({\it e.g.}, see \cite{Jia:2006rx}).
At the lowest order in $q$ and $q^\prime$, we only need retain the first term in (\ref{quark:propagator}), and neglect all the occurrences of $q^{(\prime)}$
in the numerator of the amplitude. It turns out that
Fig.~\ref{feynman:diag:gamma}$a)$ then exactly cancels against
Fig.~\ref{feynman:diag:gamma}$b)$, thus
rendering a net vanishing result at lowest order in velocity expansion~\footnote{
This is somewhat analogous to the hard exclusive
process $\eta_b\to J/\psi J/\psi$, where the amplitude also vanishes
at the lowest order in charm quark relative velocity~\cite{Jia:2006rx}.}.

To obtain a nonvanishing prediction, one must proceed to the first
order in $q^{(\prime)}$ in the amplitude. To this level of accuracy,
it is legitimate to set $E_q=E_{q^\prime} \approx m$, since the induced error is of quadratic order in $q^{(\prime)}$.
It is a curious fact that, if one still keeps only the first term in the
quark propagator (\ref{quark:propagator}), the $\mathcal {O}(q^{(\prime)})$ pieces from the
spin projectors and the $\not\!\! q^{(\prime)}$ term from the quark
propagator then make a nonzero contribution in an individual diagram,
but their contributions again cancel upon summing Fig.~\ref{feynman:diag:gamma}$a)$ and Fig.~\ref{feynman:diag:gamma}$b)$.
Therefore, the leading surviving
$\mathcal {O}(q^{(\prime)})$ contribution can only be obtained by retaining the second term
in the expanded quark propagator (\ref{quark:propagator}), while neglecting $q^{(\prime)}$ terms
altogether in elsewhere of the amplitude.
After some efforts, we can read off the reduced amplitude:
%------------------------
\bqa
%------------------------
{\mathscr A} &=&  2 {4 e e_Q g_s^2 C_F \over (k\cdot P)^2}
\int\!\!\! \int \!\! {d^3 q \over (2\pi)^3} {d^3 q^{\,\prime} \over
(2\pi)^3} \, \phi_{n^\prime 0}^\ast({\mathbf q^\prime})
%------------------------
\nn \\
%------------------------
&\times & T(\mathbf{q}^\prime-\mathbf{q}) \,
\phi_{n0}({\mathbf q}),
%------------------------
\label{factorization:momentum:space}
%------------------------
\eqa
%------------------------
where $C_F={4\over 3}$, and the prefactor 2 indicates that two
undrawn diagrams make equal contributions as
Fig.~\ref{feynman:diag:gamma}$a)+b)$, owing to charge conjugation
symmetry. $\phi_{n^{(\prime)}0}$ signifies the momentum-space
Schr\"{o}dinger wave function, and the ``hard-scattering" kernel is
%------------------------
\bqa
%------------------------
T(\mathbf{q})&=&  -{\mathbf{k}\cdot \mathbf{q} \over \mathbf{q}^2 +
\mathbf{k} \cdot \mathbf{q}- i \epsilon }.
%------------------------
\label{sh:kernel:momentum:space}
%------------------------
\eqa
%------------------------
Eq.~(\ref{factorization:momentum:space}) is the desired
factorization formula in momentum space.

It would be more convenient to work with the familiar spatial wave
functions. Thanks to the fact that the ``hard-scattering" part
depends only on the difference between the relative momenta of two quarkonia,
$\mathbf{q}^\prime-\mathbf{q}$, and not on $\mathbf{q}$ or
$\mathbf{q}^\prime$ separately, upon Fourier transformations,
one can arrive at a compact expression in the position space via contour integral:
%------------------------
\begin{subequations}
%------------------------
\bqa {\mathscr A} &=&  {4 e e_Q  C_F \alpha_s \over M_n} \,{\mathscr
E}_{n n^\prime},
%------------------------
\\
%------------------------
{\mathscr E}_{n n^\prime} &=& \int^\infty_0 \!\! d r \,r^2\,
R_{n^\prime 0}^\ast(r) \,\mathscr{T}(r) \, R_{n0}(r),
%------------------------
\eqa
%------------------------
\end{subequations}
%------------------------
where $R(r)$ appearing in the overlap integral ${\mathscr E}_{n
n^\prime}$ is the radial wave function. We have used the relation
$k\cdot P=k\,M_n$ and $M_n$ is the mass of the initial-state
quarkonium. The dimensionless kernel $\mathscr{T}(r)$ is obtained by
Fourier transforming $T(\mathbf{q})$ and integrating over solid
angle:
%------------------------
\bqa
%------------------------
\mathscr{T}(r) &=& {e^{{i \over 2}k r} \over M_n r}\left[
j_0\left({k r\over 2}\right) - {2\over k r} j_1\left({k r\over
2}\right) + i j_1\left({k r\over 2}\right) \right],
%------------------------
\nn\\
%------------------------
\eqa
%------------------------
where $j_1(x) \equiv {\sin x\over x^2}-{\cos x\over x}$.
It may be worth
reminding that the above combination of spherical Bessel functions
in the bracket resembles the conventional electric-dipole ($E1$)
transition formula with finite-size effect incorporated.
Notice $\mathscr{T}(r)$ develops an imaginary part, since the exchanged
semihard gluon can become on-shell when
$\mathbf{q}-\mathbf{q}^\prime= \mathbf{k}$.
However, we would like to stress that,
the characteristic virtuality of the exchanged
gluon in (\ref{gluon:propagator}) should be of order $m^2 v^2\gg \Lambda^2_{\rm QCD}$,
thus the emergence of imaginary part in the ``hard-scattering" kernel
should be viewed as an artifact due to ignoring the recoiling effect of ${\mathbf P}^\prime$.
If the effect of the imaginary part is insignificant with respect to that of the
real part, we may feel such a ignorance is tolerable, otherwise
it will indicate a theoretical disaster. As we will see in later phenomenological
analysis, the contamination of the imaginary part is indeed always negligible for
a class of hindered $M1$ transitions in bottomonium and charmonium systems.

It is interesting to examine the asymptotic form of $\mathscr{T}(r)$
as $kr\ll 1$. Using $j_l(x)\sim {x^l\over (2 l+1)!!}$ at small $x$,
one finds $ \mathscr{T}(r) \to {1 \over 3 m r}+i {k\over 4m}$ as
$kr\to 0$. It turns out that, the real part might be identified
with, up to a constant, the $\mathcal{O}(\alpha_s)$ matching
coefficient $V_S^{[\sigma \cdot (r\times r\times B)]/m^2}$ in
Sec.~IIIC of \cite{Brambilla:2005zw} (note there arises some subtle
issue regarding gauge invariance). Since the imaginary part becomes
$r$-independent in the long wavelength limit, as expected, it does
not contribute to the hindered $M1$ transition.

Finally we can express the transition width as
%------------------------
\bqa
%------------------------
& & \Gamma[n\,^3S_1\to n^\prime\,^1S_0+\gamma] = { k^3 \over 12\pi}
\left|{\mathscr{A}}\right|^2
%------------------------
\nn \\
%------------------------
& = & {16\over 3}\, \alpha e_Q^2  {k^3\over M_n^2} \,C_F^2\alpha_s^2
\left| {\mathscr E}_{n n^\prime}\right|^2 \,,
%------------------------
\label{final:width:3S1:1S0}
%------------------------
\eqa
%------------------------
where we have averaged upon spin of the initial ${}^3S_1$ state and
sum over two transverse polarizations of the photon.

%------------------------
\begin{table*}[ht]
 \caption{Measured and predicted branching fractions of various hindered $M1$
 transition processes $n\,^3S_1\to n^\prime\,^1S_0+\gamma$
 for bottomonium and charmonium. The photon momentum $k$ is determined by physical
 kinematics. The total widths of $\Upsilon(nS)$ and $\psi(2S)$ states, as
 well as all the quarkonium masses, are taken from PDG08
 compilation~\cite{Amsler:2008zzb}, except
 $\eta_b$ mass is taken to be 9389 MeV~\cite{Aubert:2008vj}, and
 $\eta_b(2S)$ mass taken as 9997 MeV~\cite{Godfrey:2001eb}.
 For $\Upsilon(2S){\to}\gamma\eta_b$, we use the preliminary \textsc{Babar}
 result~\cite{etabtalk:2008}; for $\psi(2S){\to}\gamma\eta_c$,
 we quote the latest \textsc{Cleo} measurement~\cite{Mitchell:2008fb},
 instead of the world average value given in \cite{Amsler:2008zzb}. We
 have taken $\alpha_s(\mu) = 0.43$ and 0.59 for $\mu=$ 1.2 and 0.9 GeV, respectively.}
 \label{3S1:to:1S0:numerical}
 \begin{ruledtabular}
 \begin{tabular}{lcccccccc}
     \multicolumn{1}{c}{Decay}&\multicolumn{1}{c}{$k$} &
     \multicolumn{1}{c}{$\mathcal B$ (Exp.)} &\multicolumn{1}{c}{$\alpha_s$}
      &\multicolumn{2}{c}{$\mathscr{E}_{n n^{\prime}}(\times10^{-2})$}&\multicolumn{2}{c}
      {$\mathcal B$ (
      Our predictions)}
      \\
     \cline{5-6}\cline{7-8}\multicolumn{1}{c}{modes} & (MeV) & & & Cornell& BT
     & Cornell & BT
     \\ \hline
%------------------------
$\Upsilon(2S){\to}\gamma\eta_b$    & 614 &
$(4.2\pm1.4)\times10^{-4}$
 & 0.43 & $3.7 e^{i2.0^\circ}$ & $3.2 e^{i2.7^\circ}$ & $1.4\times10^{-4}$&$1.1\times10^{-4}$&
 \\
%------------------------
$\Upsilon(3S){\to}\gamma\eta_b$  & 921  & $(4.8\pm
1.3)\times10^{-4}$ & 0.43 & $2.7 e^{i2.6^\circ}$ & $2.3 e^{i
3.5^\circ}$ &$3.7\times10^{-4}$ & $2.8\times10^{-4}$&
\\
%------------------------
$\Upsilon(4S){\to}\gamma\eta_b$    &1123   & -- & 0.43&$2.2
e^{i2.8^\circ}$ &$1.9 e^{i3.7^\circ}$&$4.3\times10^{-7}$ &
$3.2\times10^{-7}$&
\\
%------------------------
$\Upsilon(4S){\to}\gamma\eta_b(2S)$ & 566 & --  & 0.43 & $1.7
e^{i2.2^\circ}$ & $1.6e^{i2.7^\circ}$ & $ 3.2\times 10^{-8}$ &
$2.7\times10^{-8}$&
\\ \hline
%------------------------
$\psi(2S){\to}\gamma\eta_c$  & 638 & $(4.3 \pm 0.6)\times10^{-3}$ &
0.59 &$6.4 e^{i9.7^\circ}$ &$5.7 e^{i12.9^\circ}$
&$2.7\times10^{-3}$&$2.1 \times10^{-3}$&
\end{tabular}
 \end{ruledtabular}
 \end{table*}
%------------------------
%----------------------------------------------------------------

Eq.~(\ref{final:width:3S1:1S0}) is the key formula of this work,
which looks quite simple. In evaluating the overlap integral
${\mathscr E}_{n n^\prime}$, the input wave functions are obtained
by solving Schr\"{o}dinger equation with the widely-used
Cornell potential model~\cite{Eichten:1978tg} and Buchmuller-Tye (BT)
potential model~\cite{Buchmuller:1980su}. Parameters in both potential
models are tuned such that the $b\bar{b}$ and $c\bar{c}$
spectroscopy below open flavor threshold are successfully
reproduced.
%-------------------------------------------
The only freely adjustable parameter seems to be the strong coupling
constant, $\alpha_s(\mu)$. However, the choice of the
renormalization scale $\mu$ is by no means arbitrary. On physical ground,
it should be fixed around the typical value of the quark 3-momentum in
quarkonium, which is about $1.2$ GeV for $b\bar{b}$ system, and
$0.9$ GeV for $c\bar{c}$ system~\cite{Braaten:1996ix}. Therefore,
with $\alpha_s$ fixed, our formalism becomes rather predictive, and,
readily falsifiable.

In Table~\ref{3S1:to:1S0:numerical} we have tabulated various
predictions to hindered $M1$ transitions of $n{}^3S_1 \to
n^\prime{}^1S_0$.
%---------------------------------------
We also present the numerical results for $\mathscr{E}_{n
n^{\prime}}$, and reassuringly,
the contribution from ${\rm Im} {\mathscr T}(r)$ is indeed
insignificant.
%---------------------------------------
As one can tell, the agreement between our predictions, especially
from the Cornell potential model, and the measurement for the
transition rate of $\Upsilon(3S)\to \eta_b\gamma$, is strikingly
successful. Curiously, for other hindered $M1$ transitions, where the photon is
not that energetic so that the multipole expansion method may still
apply, our formalism again appears to make a decent account of the
measured transition rates, agrees typically within $2-3$ $\sigma$.
It seems fair to conclude that our simple factorization formula has
passed quite nontrivial tests. Given the fact that there is almost
no free parameters in (\ref{final:width:3S1:1S0}), we feel
encouraged that our formalism has captured at least some correct and
relevant ingredients. We hope future measurements of $\Upsilon(4S)
\to \eta_b \gamma$ can further test our mechanism.

It might be tempting to seek simplified expression for the overlap
integral ${\mathscr E}_{n n^\prime}$, by exploiting some hierarchy
between different $b\bar{b}$ energy levels. Higher radial
excitation, say, $\Upsilon(3S)$, is known to have considerably
larger radius than $\eta_b$. An intuitive guess is that ${\mathscr
E}_{31}$ may not be necessarily sensitive to the full profile of
$R_{30}(r)$, instead may only sensitive to its value at small
distance (about the radius of $\eta_b$). If this were true, one
could pull $R_{30}(r)$ outside of the integral, and approximate it
by its value at the origin. The transition rate predicted this way
turns out to be about two orders of magnitude greater than the
measured one! If we play the same game for $R^*_{10}(r)$, the
result would be about ten times larger than the data. The failure of
these approximations may be understood from the empirical fact that,
in the Cornell or BT potential models, the average momentum of quark
in different $b\bar{b}$ energy levels is more or less equal. As a
result, there seems no ground to neglect $\mathbf{q}$ or
$\mathbf{q}^\prime$ in the ``hard-scattering" kernel in
(\ref{factorization:momentum:space}).

For the $M1$ transition from $n{}^1S_0$ to $n^\prime{}\,^3S_1$, one
needs multiply (\ref{final:width:3S1:1S0}) by a statistical factor
of 3. Various partial widths for $\eta_b(nS)\to \Upsilon\gamma$ are
about 10 eV, and that for $\eta_c(2S)\to J/\psi\gamma$ is about 1
keV. These bottomonium transitions may be accessible in high-energy
hadron collider experiments such as CERN Large Hadron Collider
(LHC), and BESIII program may provide a chance to look for this
charmonium hindered transition.

As in any factorization framework, we expect that the
factorization formula (\ref{factorization:momentum:space}) is
perturbatively improvable. It will be a major progress to calculate
the next-to-leading order correction to the ``hard-scattering" kernel.
To achieve this, it might prove easier to reformulate our derivation
in the context of NRQCD.
It would also be interesting to implement relativistic
corrections to (\ref{factorization:momentum:space}).

Obviously our strategy needs not to be confined to hindered $M1$
transitions only. It should be applicable whenever the radiated
photon cannot be viewed as ultrasoft and multipole expansion breaks
down. It will be interesting to work out the corresponding
factorization formula for $E1$ transitions such as
$\chi_{bJ}(2P)\to \Upsilon\gamma$. It would be also interesting to
generalize this ``hard-scattering" formalism to explore the hadronic transition
processes such as $\Upsilon(3S,4S)\to \Upsilon+\pi\pi$.
\begin{acknowledgments}
%----------------------------------------------------------------
We thank Antonio Vairo, Wei Wang and Yu-Ming Wang for useful
discussions.
%on their unpublished results about the process $\Upsilon(3S)\to
%\eta_b\gamma$ by utilizing different theoretical approaches.
%----------------------------------------------------------------
This research was supported in part by the National Natural Science
Foundation of China under grants No.~10875130, 10605031, and 10935012.
\end{acknowledgments}

%%%%%%%%%%%%%%%%%%%%%%%%%%%%%%%%%%%%%%%%%%%%%%%%%%%%%%%%%%%%%%%%%%%%%%%%%%%%%%
\end{document}